\documentclass{elsart}

\usepackage{cite}

\usepackage{dcolumn}

\usepackage{graphicx}

\begin{document}

\begin{frontmatter}

\title{Comment on: ``Threading dislocation densities in semiconductor crystals: A
geometric approach''. Phys. Lett. A 376 (2012) 2838-2841}
\author{Francisco M. Fern\'{a}ndez \thanksref{FMF}}

\address{INIFTA, DQT, Sucursal 4, C.C 16, \\
1900 La Plata, Argentina}

\thanks[FMF]{e--mail: fernande@quimica.unlp.edu.ar}

\begin{abstract}
We analyze the solution of the Schr\"odinger equation arising in the
treatment of a geometric model introduced to explain the origin of the
observed shallow levels in semiconductors threaded by a dislocation density.
We show (contrary to what the authors claimed) that the model does not
support bound states for any chosen set of model parameters. Assuming a
fictitious motion in the $x-y$ plane there are bound states provided that $%
k\neq 0$ and not only for $k>0$ as the authors believed. The truncation
condition proposed by the authors yields only one particular energy for a
given value of a chosen model parameter and misses all the others
(conditionally solvable problem)
\end{abstract}


\end{frontmatter}

Some time ago Bakke and Moraes\cite{BM12} introduced a geometric model to
explain the origin of the observed shallow levels in semiconductors threaded
by a dislocation density. It leads to a Schr\"{o}dinger-Pauli equation that
can be reduced to a biconfluent Heun equation by means of suitable
transformations. Through application of the Frobenius method the authors
derived a three-term recurrence relation for the expansion coefficients.
They claimed that it was necessary to resort to a truncation condition in
order to have bound states and in this way they obtained analytical
expressions for the energies of the model. The purpose of this Comment is
the analysis of the procedure proposed by Bakke and Moraes\cite{BM12} for
solving their eigenvalue equation.

The starting point is the time-dependent Schr\"{o}dinger-Pauli equation
\begin{eqnarray}
i\frac{\partial \psi }{\partial t} &=&-\frac{1}{2m}\left[ \frac{\partial ^{2}%
}{\partial \rho ^{2}}+\frac{1}{\rho }\frac{\partial }{\rho }+\frac{1}{\rho
^{2}}\frac{\partial ^{2}}{\partial \varphi ^{2}}-2\Omega \frac{\partial ^{2}%
}{\partial \varphi \partial z}+\left( 1+\Omega ^{2}\rho ^{2}\right) \frac{%
\partial ^{2}}{\partial z^{2}}\right] \psi +  \nonumber \\
&&\frac{i\sigma ^{3}}{2m\rho ^{2}}\frac{\partial \psi }{\partial \varphi }-%
\frac{i\Omega \sigma ^{3}}{2m}\frac{\partial \psi }{\partial z}+\frac{1}{%
8m\rho ^{2}}\psi +\frac{\Omega ^{2}}{8m}\psi +\frac{f}{\rho }\psi ,
\label{eq:time-dependent_Schro}
\end{eqnarray}
where the meaning of the parameters can be seen in the authors' paper. The
form of the Laplacian operator $\nabla ^{2}$ comes from the line element
\begin{equation}
ds^{2}=d\rho ^{2}+\rho ^{2}d\varphi ^{2}+\left( dz+\Omega \rho ^{2}d\varphi
\right) ^{2},  \label{eq:line_element}
\end{equation}
in cylindrical coordinates $0\leq \rho <\infty $, $0\leq \varphi \leq 2\pi $%
, $-\infty <z<\infty $. The authors argue as follows ``We can see in Eq. (7)
that $\psi $ is an eigenfunction of $\sigma ^{3}$, whose eigenvalues are $%
s=\pm 1$ and the Hamiltonian of Eq. (7) commutes with the operators $\hat{J}%
_{z}=-i\partial _{\varphi }$ and $\hat{p}_{z}=-i\partial _{z}$, thus, we can
write the solution of Eq. (7) in terms of the eigenfunctions of the
operators $\hat{J}_{z}$ and $\hat{p}_{z}$, that is, $\psi _{s}=e^{-i\mathcal{%
E}t}e^{i\left( l+\frac{1}{2}\right) \varphi }e^{ikz}R_{s}(\rho )$, where $%
l=0,\pm 1,\pm 2,...$ and $k$ is a constant which corresponds to the momentum
in the $z$-direction. We take $k>0$ since, as we will see below in Eq. (10),
for $k<0$ the minus sign of the exponent of the Gaussian function becomes
positive and we no longer have bound states. This asymmetry is due to the
choice of the Burgers vector orientation.'' The statement about the allowed
values of $k$ is wrong as we shall see in what follows. To begin with $\psi
_{s}(t,\rho ,\varphi ,z)$ is a bound state only if
\begin{equation}
\int \int \int \left| \psi _{s}(t,\rho ,\varphi ,z)\right| ^{2}\rho d\rho
d\varphi dz<\infty ,  \label{eq:bound-state_def}
\end{equation}
as shown in any textbook on quantum mechanics\cite{CDL77}. In the example
discussed here the improper integral over $z$ is obviously divergent and,
consequently, there are no bound states for any chosen set of values of the
model parameters.

The function $R_{s}(\rho )$ satisfies the eigenvalue equation
\begin{eqnarray}
&&-\frac{1}{2m}\left( \frac{\partial ^{2}R_{s}}{\partial \rho ^{2}}+\frac{1}{%
\rho }\frac{\partial R_{s}}{\rho }\right) +\frac{\gamma _{s}^{2}}{2m\rho ^{2}%
}R_{s}-\frac{\Omega k\gamma _{s}}{m}R_{s}+\frac{1}{2m}\left( k+s\frac{\Omega
}{2}\right) ^{2}R_{s}+  \nonumber \\
&&\Omega ^{2}k^{2}\rho ^{2}R_{s}+\frac{f}{\rho }R_{s}=\mathcal{E}R_{s},
\label{eq.eigeval_eq_1}
\end{eqnarray}
where $\gamma _{s}=l+\frac{1}{2}(1-s)$. The authors claimed to have obtained
the equation
\begin{equation}
\frac{d^{2}R_{s}}{d\zeta ^{2}}+\frac{1}{\zeta }\frac{dR_{s}}{d\zeta }-\frac{%
\gamma _{s}^{2}}{\zeta ^{2}}R_{s}-\zeta ^{2}R_{s}-\frac{f^{\prime }}{\sqrt{%
\Omega k}\zeta }R_{s}+\frac{\beta _{s}}{\Omega k}R_{s}=0,
\label{eq.eigeval_eq_2}
\end{equation}
by means of the change of variables $\zeta =\sqrt{\Omega k}\rho $. Here the
authors made two mistakes, the first one coming from ignoring the fact that
they multiplied the whole equation by $2m$; therefore, the resulting
harmonic term should be $-2m\zeta ^{2}R_{s}$ instead of the fourth term of
this equation. The correct transformation that leads to that fourth term is $%
\zeta =\left( 2m\Omega ^{2}k^{2}\right) ^{1/4}\rho =(2m)^{1/4}\sqrt{|\Omega
k|}\rho $, provided that $k\neq 0$ (see a recent pedagogical article on
deriving dimensionless equations\cite{F20}). The second mistake is that $%
\sqrt{k^{2}}=|k|$ (and not $k$) so that the sign of $k$ does not affect the
form of the Gaussian function as the authors believed. The result of these
mistakes is that most of the parameters in equation (\ref{eq.eigeval_eq_2})
are wrong; however, their precise form is not relevant for the following
discussion.

By means of the transformation
\begin{equation}
R_{s}(\zeta )=\zeta ^{\left| \gamma _{s}\right| }e^{-\frac{\zeta ^{2}}{2}%
}\sum_{j=0}^{\infty }a_{j}\zeta ^{j},  \label{eq:R_series}
\end{equation}
the authors derived the three-term recurrence relation
\begin{eqnarray}
a_{m+2} &=&\frac{b}{(m+2)(m+\alpha +1)}a_{m+1}+\frac{2m-g}{(m+2)(m+\alpha +1)%
}a_{m},  \nonumber \\
m &=&-1,0,1,2\ldots ,\;a_{-1}=0,\;a_{0}=1,  \label{eq:rec_rel}
\end{eqnarray}
where $\alpha =2\left| \gamma _{s}\right| +1$. The parameters $b$ and $g$
are given in the authors' paper but their form is incorrect because of the
unsuitable change of variables discussed above. However, their precise form
is not relevant for the following discussion (as stated above).

The authors reasoned that ``Note that the recurrence relation (14) is valid
for both signs of the Coulomb-like potential, that is, we can specify the
signs of the Coulomb-like potential by making $f\rightarrow \pm |f|$ in
(14). Hence, in order to obtain finite solutions everywhere, which represent
bound state solutions, we need that the power series expansion (13) or the
Heun biconfluent series become a polynomial of degree $n$. Through
expression (14), we can see that the power series expansion (13) becomes a
polynomial of degree $n$ if we impose the conditions:
\begin{equation}
g=2n\ \mathrm{and}\ a_{n+1}=0,  \label{eq:trunc_cond}
\end{equation}
where $n=1,2,3,\ldots $.'' From the first truncation condition the authors
derived an analytical expression for the energy $\mathcal{E}_{n,l,s}$ and
stated that ``where the angular frequency is given by $\omega _{n,l,s}=\frac{%
\Omega k}{m}$. On the other hand, the condition $a_{n+1}=0$ allows us to
obtain an expression involving the angular frequency and the quantum numbers
$n$, $l$ and $s$. Observe that we have considered $k$ being a positive
constant, thus, we can choose any value of $k$ in such a way that the
condition $a_{n+1}=0$ and write $k=k_{n,l,s}$. We should note that writing $%
k $ in terms of the quantum numbers $n$, $l$ and $s$, that is, $k=k_{n,l,s}$%
, it does not mean that $k$ is quantized. Writing $k=k_{n,l,s}$ means that
the choice of the values of $k>0$ depends on the quantum numbers $n$, $l$
and $s$ in order to satisfy the condition $a_{n+1}=0$. Then, we can write
the angular frequency as $\omega _{n,l,s}=\frac{\Omega k_{n,l,s}}{m}$.'' The
authors appear to believe that there are bound states only when the
truncation conditions (\ref{eq:trunc_cond}) are satisfied and only for $k>0$%
. Besides, they said that the truncation conditions (\ref{eq:trunc_cond})
require that $k$ take some particular values $k=k_{n,l,s}>0$ without being
understood as a quantization of $k$. These statements are wrong as we show
in what follows (we have already shown above that the only restriction is
that $k\neq 0$).

First of all we want to make it clear that the model chosen by the authors
does not support bound states as argued above. For this reason, in what
follows we assume that the eigenvalue equation (\ref{eq.eigeval_eq_1})
applies only to the motion of a particle in the $x-y$ plane (with $k\neq 0$%
!) so that we can speak of bound states. In such a case we have bound states
for all values of $f$ if $k\neq 0$ and for $f<0$ when $k=0$. It is plain
that it is not necessary to restrict the analysis only to positive values of
$k$.

In order to make the analysis simpler we rewrite equation (\ref
{eq.eigeval_eq_2}) as follows
\begin{equation}
\frac{d^{2}R_{s}}{d\zeta ^{2}}+\frac{1}{\zeta }\frac{dR_{s}}{d\zeta }-\frac{%
\gamma _{s}^{2}}{\zeta ^{2}}R_{s}-\zeta ^{2}R_{s}-\frac{b}{\zeta }%
R_{s}+WR_{s}=0,  \label{eq.eigeval_eq_3}
\end{equation}
that is obviously valid only for $k\neq 0$. As stated above there are bound
states for all values of $b$ because the harmonic term determines the
behaviour at infinity and $\gamma _{s}^{2}/\zeta ^{2}$ the behaviour at
origin. Any textbook on quantum mechanics\cite{CDL77} shows that bound
states occur for those values of $W$ such that
\begin{equation}
\int_{0}^{\infty }\left| R_{s}(\zeta )\right| ^{2}\zeta \,d\zeta <\infty .
\label{eq:bound-state_def_radial}
\end{equation}
On the other hand, the truncation condition (\ref{eq:trunc_cond}) yields
bound states for particular values of $b=b_{n,l,s}$ and for each of them
they provide just one eigenvalue $W_{n,l,s}$. Equation (\ref{eq.eigeval_eq_3}%
) leads to the three-term recurrence relation (\ref{eq:rec_rel}) with $%
g=W-2|\gamma _{s}|-2$; so that from the truncation conditions (\ref
{eq:trunc_cond}) for $n=1$ we obtain $b=\pm \sqrt{\alpha g}$. Therefore,
when $s=1$ and $l=1$ we have $\gamma _{1}=1$, $g=2$, $b=\pm \sqrt{6}$ and $%
W=6$.

Let us solve the eigenvalue equation (\ref{eq.eigeval_eq_3}) by means of
standard approximate methods; for example, the reliable Rayleigh-Ritz
variational method that is known to yield upper bounds to all states\cite
{P68}. For simplicity we choose the basis set of (un-normalized) functions $%
\left\{ u_{j}(\zeta )=\zeta ^{|\gamma _{s}|+j}e^{-\frac{\zeta ^{2}}{2}%
},\;j=0,1,\ldots \right\} $. In order to test the accuracy of the
variational results we resort to the powerful Riccati-Pad\'{e}
method\cite {FMT89a}. For $b=\sqrt{6}$ and $\gamma _{1}=1$ the
three lowest eigenvalues given by both methods are $W_{0,1,1}=1.600357154$, $W_{1,1,1}=6$%
, $W_{2,1,1}=10.21072810$. Since the differential equation (\ref
{eq.eigeval_eq_3}) is invariant under the transformation $(b,\zeta
)\rightarrow (-b,-\zeta )$ we conclude that $W(-b)=W(b)$ and the numerical
calculations just mentioned with $b=-\sqrt{6}$ confirm this general result.
The important point here is that the method proposed by Bakke and Moraes\cite
{BM12} only yields one eigenvalue for a particular value $b_{n,l,s}$ of the
model parameter $b$ and misses all the other ones. This fact is not
surprising as it is expected of the exact solutions to
conditionally-solvable models\cite{D88, BCD17} (and references therein) and
is confirmed by the numerical calculation discussed above. It is obvious
that each of the energies $\mathcal{E}_{n,l,s}$ reported by those authors
correspond to a particular model with the parameter $b_{n,l,s}$ (of course
if we omit the fact that almost all the relevant parameters, energy
included, appearing in the equations below their equation (8) are wrong).

Summarizing: the model proposed by Bakke and Moraes\cite{BM12}
does not support bound states (for any chosen set of model
parameters) in contradiction with what they claimed. Assuming a
fictitious motion in the $x-y$ plane there are bound states
provided that $k\neq 0$ and not only for $k>0$ as they stated. The
truncation condition proposed by the authors yields only one
particular energy for a given value of a chosen model parameter
and misses all the other eigenvalues (conditionally solvable
problem).

\end{document}